\documentstyle [fleqn,12pt]{article}
\newcommand{\dy}[2]{\frac{{\textstyle \partial \/} #1}{{\textstyle 
\partial \/} #2}}
\newcommand{\df}[2]{\frac{{\textstyle d \/} #1}{{\textstyle d \/} #2}}
\newcommand{\ab}[1]{{| \/ } #1 |}
\begin{document}
\baselineskip .8cm 
\begin{titlepage}
\title{ \bf Thermalization in Yang-Mills Classical Mechanics }      
\author{Vishnu M. Bannur \\
{\it Department of Physics}, \\  
{\it University of Calicut, Kerala-673 635, India.} }
\date{}
\maketitle
\begin{abstract} 
	We use quartic oscillators system with two degrees of freedom 
to model Yang-Mills classical mechanics. This simple model explains 
qualitatively many features reported in lattice calculation of 
$(3+1)$ - dimensional classical Yang-Mills system. 
The largest Lyapunov exponent ($\lambda$) and the thermalization 
time were numerically evaluated. We also show, in our model, 
that $\lambda$ scales with 4th root of energy density. Here thermalization 
is due to relaxation phenomena associated with the color degrees of freedom. 
From the physical picture of thermalization, we speculate that the 
system with coherent fields (flux tubes) formed in relativistic heavy ion 
collisions can relax by chaos and the estimated thermalization time can be  
smaller than $1 \; fm/c$. 
\end{abstract}
\vspace{1cm}

\end{titlepage}
  Quantum chromodynamics (QCD) is a Yang-Mills (YM) theory,   
  nonlinear and hence exhibits chaos \cite{sa.1,bi.1}. 
  This property is used to understand the 
  confinement as well as deconfinement problems in QCD. 
  Savvidy \cite{sa.1} extensively studied chaos in 
a homogeneous $SU(2)$ Yang-Mills system and speculated that it 
might be related to the confinement problem. Recently, M\"uller {\it et al.} 
\cite{mu.1} studied the classical YM system in $(3+1)$ - dimension on 
a lattice and showed that it exhibits deterministic chaos. They related  
YM chaos to the thermalization of gluons in the deconfined state, 
 which may be formed in relativistic heavy ion collisions (RHICs). 
 They evaluated the maximum Lyapunov exponent, $\lambda$, and found that 
 it scales with energy per plaquette and  
 estimated the thermalization time of QGP formed in RHICs. 
 But their analysis involves heavy numerical 
 computations. It was also not clear why $\lambda$ linearly depends on energy 
per plaquette. 
 In the light of their results, it is interesting to 
 reanalyse the earlier work on Yang-Mills classical mechanics (YMCM), 
 studied by Savvidy.  

 In this letter we study this problem using a 
 very simple model, quartic oscillators (QO), which is related to YMCM of 
 $SU(2)$ under certain limit. The Lyapunov exponent 
 (LE) of QO was studied by 
 Joy and Sabir \cite{jo.1} and it's statistical mechanics and 
 thermodynamics was studied by us \cite{ba.1}. 
 Here we find that maximum LE, $\lambda$, scales with 4th root of energy 
 density, which can be seen both by a dimensional argument  
 as well as numerically and thermalization time is estimated. 
 This is similar to the result of M\"uller {\it et al.}, but 
 obtained with a simpler model, less numerical computations and  
with a clear physical picture of thermalization based on the 
statistical mechanics of chaotic system. The coherent field energy, like 
flux tubes formed in relativistic heavy ion collisions, may be  
redistributed among various degrees of freedom associated with spatial, 
color, chromo electric and magnetic fields due to chaos in YM theory. 

The YMCM Hamiltonian is a $SU(2)$ YM Hamiltonian density in temperal gauge, 
$A_0^a = 0$, and assuming that the vector potentials $\vec{A}^a (t)$ 
are functions of time only. It is given by 
\begin{equation} H_{YM} = \frac{1}{2} \sum_a \dot{\vec{A}}_a^2  
              + \frac{1}{4}\, g^2\, \sum_{a,b} 
	     (  \vec{A}_a \times \vec{A}_b )^2 \; , \label{eq:ym} \end{equation} 
where $g$ is the gauge coupling constant and $a=1,2,3$ are color quantum 
numbers. The Hamiltonian, $H_{YM}$, may be rewritten in the form 
 \begin{equation} H_{YM} = H_{FS} + T_{YM} 
	     = \frac{1}{2} (\dot{x}^2 + \dot{y}^2 + \dot{z}^2)   
	       + \frac{1}{2} g^2 (x^2 y^2 + y^2 z^2 + z^2 x^2)   
	       + T_{YM} \; , \end{equation} 
where $H_{FS}$ is called a fundamental subsystem (FS) of YMCM and 
$T_{YM}$ describes quasi-rotational freedoms. Let us consider a
simpler two dimensional model ($a=1,2$ in Eq. (\ref{eq:ym})) and 
also without $T_{YM}$. The corresponding Hamiltonian is 
 \begin{equation} H_{2} = \frac{1}{2} (\dot{x}^2 + \dot{y}^2)   
	       + \frac{1}{2} g^2 x^2 y^2 \; . \label{eq:2ym} \end{equation} 
On redefining variables $X_1 \equiv g\, x$ and $X_2 \equiv g\, y$, it reduces to  
 \begin{equation} H \equiv g^2 H_2 = \frac{1}{2} (\dot{X}_1^2 + \dot{X}_2^2)   
	       + \frac{1}{2} X_1^2 X_2^2 \; . \label{eq:ym2} \end{equation} 
 It is very similar to the quartic oscillator (QO) system with two 
 degrees of freedom, which has been studied extensively both  
 classicaly and quantum mechanicaly and is given by \cite{jo.1} 
 \begin{equation} H_Q = \frac{1}{2} (\dot{X}_1^2 + \dot{X}_2^2)  
	       + \frac{(1-\alpha)}{12} (X_1^4 + X_2^4)  
	       + \frac{1}{2} X_1^2 X_2^2 \; , \label{eq:qo} \end{equation} 
 where $\alpha$ is a parameter. For $\alpha=1$ it reduces to Eq.(\ref{eq:ym2}). 
 QO  system is highly chaotic for $\alpha=1$ and becomes less chaotic 
 as $\alpha$ decreases as shown in Ref. \cite{jo.1}. 

 When we make a connection between YMCM and QO, we may note that 
 the variables $X_i$ have dimensions of energy ($E$). $H_Q$ is a 
 constant of motion, say $g^2 \varepsilon$, and has dimension of energy density. 
 Hence we normalize all variables by 4th root of $g^2 \varepsilon$ (say, $E$) 
 to make them dimensionless. Thus we get, 
 \begin{equation} H' = \frac{1}{2} (\dot{q}_1^2 + \dot{q}_2^2)  
	       + \frac{(1-\alpha)}{12} (q_1^4 + q_2^4)  
	       + \frac{1}{2} q_1^2 q_2^2 \; , \end{equation} 
 where $H' = 1$, $q_i \equiv X_i/E$ and $\dot{q}_i \equiv 
 \df{\textstyle q_i}{\textstyle \tau}$ and     
 $\tau \equiv E\, t$. Now we can define the distance between two 
 trajectories, $D(\tau)$, in phase space, as  
\begin{equation} D^G (\tau) = \sqrt{ \sum_{i=1}^{2} \left( (q_i - q'_i)^2 
		  + (\dot{q}_i - \dot{q'}_i)^2 \right) } \; , \end{equation} 
 where primed and unprimed variables describe two different trajectories. 
 Superscript $G$ refers to the fact that it is the general definition 
 used in text books \cite{tx.1} and Ref. \cite{jo.1}. Note that without 
 normalized variables, $D(t)$ is not dimensionally correct. Now the 
 maximum Lyapunov exponent is defined as 
 \begin{equation} \lambda_1 \equiv \lim_{\tau \rightarrow \infty} \, \frac{1}{\tau} 
            \ln \ab{\frac{D(\tau)}{D(0)} } = \frac{\lambda_{E}}{E}  \; , \end{equation} 
 where 
 \begin{equation} \lambda_{E} \equiv \lim_{t \rightarrow \infty} \, \frac{1}{t} 
            \ln \ab{\frac{D(t,E)}{D(0,E)} } \; . \label{eq:lt} \end{equation} 
 Here $D(t,E)$ is given by  
\begin{equation} D(t,E) = \frac{1}{E}\, \sqrt{ \sum_{i=1}^{2} \left( (X_i - X'_i)^2 
  + (\dot{X}_i - \dot{X'}_i)^2 / E^2 \right) } \;\; \; , \label{eq:dt} \end{equation} 
 $\lambda_{E}$ is equal to $\lambda_1$ for $E=1$ and for arbitrary $E$ or 
 $g^2 \varepsilon$, $\lambda_E$ scales with $E$ or the 4th root of 
 $\varepsilon$ which has the dimension of energy. This is similar to the results 
 of M\"uller {\it et al.}, obtained numerically, whereas here  
 it follows from a dimensional arguments. We have also numerically 
 verified Eq. (\ref{eq:lt}) with our model; this is tabulated in 
 Table 1.  

 The maximum Lyapunov exponent is evaluated using the procedure of 
 Joy and Sabir \cite{jo.1} using the Hamiltonian, Eq. (\ref{eq:qo}), with 
 $\alpha = 0.99$, which is close to 1. For $\alpha$ exactly equal to 1, numerical 
 calculations are not reliable and large numerical error develops 
 which can be seen from the numerical values of $\varepsilon$, which is 
 a constant of motion. The Hamilton equations of motion, to be solved 
 numerically, are 
\[ \dot{X}_1 = X_3 \; ; \; \dot{X}_3 = (\alpha - 1) X_1^3 / 3 - X_1 X_2^2 \;;\]   
\begin{equation} \dot{X}_2 = X_4 \; ; \;  
  \dot{X}_4 = (\alpha - 1) X_2^3 / 3 - X_2 X_1^2 \; , \label{eq:qo1} \end{equation} 
 which follow from the Eq. (\ref{eq:qo}). 
 These equations are to be evolved along with the equation for 
 their variations $Y_i \equiv \delta X_i $, which are given by 
 \[ \dot{Y}_1 = Y_3 \; ; \; 
 \dot{Y}_3 = ( (\alpha - 1) X_1^2 - X_2^2) Y_1 - 2 X_1 X_2 Y_2 )  
  \; ; \] 
 \begin{equation}  \; \dot{Y}_2 = Y_4 \; ; \;  
 \dot{Y}_4 = ( (\alpha - 1) X_2^2 - X_1^2) Y_2 - 2 X_1 X_2 Y_1 )\; . 
 \label{eq:dqo} \end{equation} 
Then $\lambda_E$ is evaluated using the Eq. (\ref{eq:lt}) for 
different $E$ and is tabulated 
in Table 1. It is in good agreement with the scaling relation. 

However, it should be noted that the procedure used by M\"uller to 
evaluate $\lambda$ is not reliable here. They obtained $\lambda$ from the slope of 
the log-plot of $D(t)$ $Vs$ $t$. But here the slope depends on 
the initial conditions one uses. 
As an example a plot is given in Fig. 1, 
where two trajectories, separated initially by small distance,
are evolved using the Eq. (\ref{eq:qo1}) for $\alpha = 0.99$ with 
different initial conditions but with same energy ($E = 1$). 

Generally in gases and liquids, thermalization is 
 due to collisions and the ratio of Kolmogorov entropy (KS-entropy) to 
 collison frequency is related to the change in thermodynamic 
 entropy $S$ \cite{mi.1}. In our case there are no 
 collisions but the nonlinearity of Yang-Mills equations drives the 
 system chaotic or ergotic and hence the thermalization. 
 KS-entropy is related to the sum of positive Lyapunov 
 exponents and here we have only one positive LE and hence $S_{KS} = \lambda_E$. 
 Let $t_{th}$ is the thermalization time, then $S_{KS} t_{th} \approx S$ or 
\begin{equation} t_{th} \approx S/S_{KS} = S(E)/\lambda_E \propto \frac{\textstyle 1}{\textstyle \lambda_1 E} \;. \end{equation}   
If we assume that our system is a subsystem and is in equilibrium with a
larger system with average energy density ($\varepsilon$) propotional to the 
$4th$ power of temperature ($T$), then $t_{th}$ is inversely propotional 
to $T$; this is consistent with earlier $(3+1)$ - dimensional calculations 
\cite{bi.1,mu.1}. In addition there is a logarithmic dependence on $T$ 
because of $S$, which depends on energy density 
($S \propto \log (\varepsilon)$) as discussed in Ref. \cite{ba.1}. 

$(3+1)$ - dimensional YM system, discussed in Ref. \cite{bi.1,mu.1}, 
is a system with large degrees of freedom due to spatial 
dependence, and the thermalization there occurs, due to equal energy 
sharing between various modes by nonlinear interactions. 
In our case, we have Yang-Mills 
fields at a point and energy sharing is due to their nonlinear 
interactions between only 2 degrees of freedom. It should be noted that 
recent studies of systems with fewer degrees of freedom (even two degrees 
of freedom), which are chaotic systems, do exhibits thermalization or equipartitioning 
of energy and  other statistical properties as discussed in Ref. \cite{ba.1} 
and reference there in. Here it is chaos which plays the role of collisions 
in statistical mechanics. In statistical mechanics, indeed we need large 
degrees of freedom to have a chaotic or ergotic motion. But here, even two 
degrees of freedom exhibits chaos and hence almost ergotic because of 
nonlinearity in the system. To understand it further,  
we may rewrite the Hamiltonian, Eq. (\ref{eq:2ym}), as 
 \begin{equation} H_{2} = \frac{1}{2} (E_1^2 + E_2^2)   
	       + \frac{1}{2} B_3^2 \; ,  \end{equation} 
using the defintion of electric and magnetic fields in terms 
of field tensor 
\[ G^{\mu \nu}_{a} = \partial^{\mu} A^{\nu}_{a} - \partial^{\nu} A^{\mu}_{a}
+ g \epsilon_{abc} A^{\mu}_{b} A^{\nu}_{c} \;, \]
$a,\;b,\;c$ are color indices
which take values 1, 2, 3 and Lorentz indices $\mu, \nu\;=0, 1, 2, 3$ with
metric $(1,-1,-1,-1)$. $\epsilon_{abc}$ is antisymmetric Levi-Civita tensor.
In our case $a=1,2$ and we use hedge hog ansatz and hence we have 
electric fields $E_1$, $E_2$ and magnetic field $B_3$. Now thermalization 
means sharing energy between electric and magnetic fields. If we 
start with whole energy, say, in electric field $E_1$, after thermalization 
the energy will be equally distributed between $E_1\;,E_2$ and $B_3$. As 
we discussed in \cite{ba.1}, whenever a system is almost chaotic 
equipartition of energy takes place and  
\begin{equation} \left< \dot{x}_1\,\dy{H_2}{\dot{x}_1} \right> 
 = \left< \dot{x}_2\,\dy{H_2}{\dot{x}_2} \right> 
 = \left< x_1\,\dy{H_2}{x_1} \right> 
 = \left< x_2\,\dy{H_2}{x_2} \right> \; , \end{equation}  
which is same as  
\begin{equation} \left< E_1^2 \right> = \left< E_2^2 \right> 
= \left< B_3^2 \right> = \frac{2}{3} \varepsilon \; , \end{equation}  
Here the angular brackets indicate the time average, which
 is also equal to the phase space average by ergotic theorem. 
 Hence $t_{th}$ gives the time required to equilibrate energy 
 between electric and magnetic fields and their components. 
 This is a very important property of YM fields which, for example  
 in relativistic heavy ion collisions, will thermalize 
 the coherent fields formed immediately after the collision.  
 This is due to the relaxation phenomena associated with color degrees 
 of freedom rather than momentum relaxation. Further more, if we extrapolate 
 the results of Ref. \cite{ba.1} on the quartic oscillators, with large 
 degrees of freedom, to our system with large color degrees of freedom or 
 large number of YMCM systems with few color degrees of freedom, the energy 
 distribution in any one component of fields may be exponential 
 decay with decay constant inversely proportional to temperature. 
In other words, corresponding field distribution is Gaussian with 
it's width proportional to temperature. 
This is similar to the case of momentum relaxation, 
 where one gets Gaussian momentum distribution of particles in 
statistical mechanics. It implies that, at low temperature, the 
probability to find any field component with low energy is high and  
hence the large number of gluons are with low energy. 
That is, low energy gluons macroscopically (largely) occupy YM system 
at low temperature which may be related to confinement. But at high 
temperature, field components with higher energy increase, but those  
of low energy decrease. Finally the probability to find low energy 
and high energy gluons will be of the same order. Now YM system may be 
in deconfined state.    

So far our discussion is based on $\lambda$ which is evaluated using 
Eq. (\ref{eq:dt}). However, in Ref. \cite{mu.1}, they used somewhat 
different definition for $D(t)$. They defined 
\begin{equation} D^M (t) = \ab{\sum_i (E_i^2 - E^{'2}_i)} \; , 
\end{equation} 
where $E_i$ are the electric field components, which reduces 
(in our notations) to 
\begin{equation} D^M (t) = \ab{\sum_i (\dot{X}_i^2 - \dot{X'}_i^2)}/g = 
		2 \ab{\sum_i \dot{Y}_i \dot{X}_i} /g \; . \label{eq:dm} \end{equation} 
$\dot{X}_i$ and $\dot{Y}_i$ are evaluated from the Eqs. (\ref{eq:qo1}) and 
(\ref{eq:dqo}). Even though $D^M (t)$ is different from $D^G (t)$, we 
get the same results. If we take square root instead of modulus 
in Eq. (\ref{eq:dm}) we get $\lambda_E$ about half of $\lambda_E$ obtained 
from Eq. (\ref{eq:dt}). This is clear from the expression for $D^M (t)$ 
where the exponentially diverging term $Y_i$ appears as linear term, 
whereas in $D^G (t)$ it is quadratic.  

In our numerical calculations, as can be seen from the Table 1, 
theoretical scaling behaviour of $\lambda$ with $E$ is confirmed. 
We also point out that the evaluation of $\lambda$ by the slope 
measurement of log plot of $D(t)$ $Vs$ $t$, where $D(t)$ is evaluted 
from two independent nearby trajectories, is not reliable. Where as 
standard method given in text books gives reliable results. This is 
probably because of the fact that LE concept is at the linear 
perturbation level, Eq. (\ref{eq:dqo}), and not with full nonlinear 
variations, as in Ref. \cite{mu.1} (by following two nearby 
trajectories separately). 

In conclusion, using a simple QO model to represent the homogeneous $SU(2)$ 
YM fields, many qualitative properties obtained by extensive numerical 
calculation of $(3+1)$ - dimensional YM fields can be reproduced. 
We obtained the linear relationship between maximum Lyapunov exponent 
and the 4th root of energy density, E, numerically as well as from a dimensional 
arguments.
The slope of the linear relationship is $0.4$ which is of the same order 
as that of M\"uller {\it et al.}. 
We also obtained an expression for thermalization time and explained the 
physical picture of thermalization in our model. The thermalization time 
is of the order of $1/(\lambda_1 E) \approx 1 fm/c$ for $g = 1$ and the energy 
density $\varepsilon = 1 \; GeV/ fm^3$ and inversely depends on the $4th$ power 
of energy density. In the case of relativistic heavy ion collisions,   
it may give an estimate of the order of magnitude of time required to 
redistribute the energy, 
stored initially in coherent fields, among all degrees of 
freedom associated with spatial, color, electric and magnetic fields, etc. 

\newpage
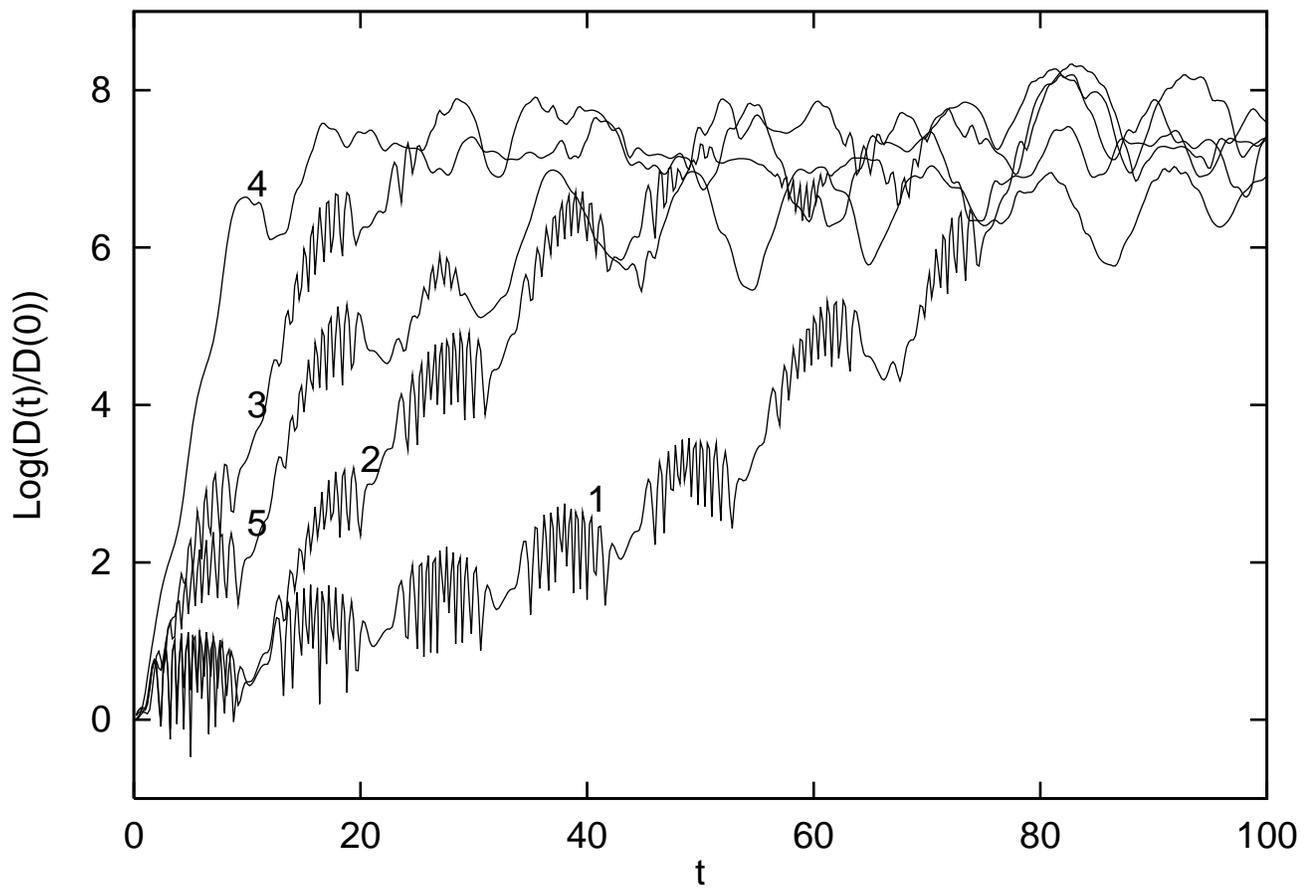
\begin{figure}
\begin{center} 
{\bf Figure Caption}\\[.5cm]  
\caption { Plots of $\ln \ab{\frac{D(t,E)}{D(0,E)} }$  as a function of $t $ 
 for $\alpha = 0.99$, $E=1$ with initial conditions $X_3$ = .01 (curve 1), 
 0.1 (curve 2), 0.4 (curve 3), 1.0 (curve 4), 1.393 (curve 5) respectively 
 with all other $X_i$ = 0.}  
\label{fig 1}
\end{center} 
\end{figure} 
\newpage 
\begin{center}
{\bf Table 1} \\[1cm]  
\begin{tabular}{||c|c|c|c|c|c||}
\hline
$g^2 \varepsilon$ &1 &16 &81 &256 &625 \\  
\hline
$\lambda_E$ &0.4 &0.78 &1.17 &1.56 &1.95 \\  
\hline
\end{tabular}  
\end{center}
\end{document}